# Nanoscale thermal control of a single living cell enabled by diamond heater-thermometer

**Alexey M. Romshin[1,*], Vadim Zeeb[2,*], Evgenii Glushkov[3], Aleksandra Radenovic[3], Andrey G. Sinogeikin[4], Igor I. Vlasov[1]**

[1]Prokhorov General Physics Institute of the Russian Academy of Sciences, Vavilov str. 38, Moscow, Russia 119991.

[2]Institute of Theoretical and Experimental Biophysics of the Russian Academy of Sciences, Pushchino, Moscow Region, Russia 142292.

[3]Laboratory of Nanoscale Biology, Institute of Bioengineering, Ecole Polytechnique Federale de Lausanne (EPFL), CH-1015 Lausanne, Switzerland

[4]Wonder Technologies LLC, Moscow, Russia.

Corresponding authors: alex_31r@mail.ru, zeebvad@gmail.com

## Abstract

We report a new approach to controllable thermal stimulation of a single living cell and its compartments. The technique is based on the use of a single polycrystalline diamond particle containing silicon-vacancy (SiV) color centers. Due to the presence of amorphous carbon at its intercrystalline boundaries, such a particle is an efficient light absorber and becomes a local heat source when illuminated by a laser. Furthermore, the temperature of such a local heater is tracked by the spectral shift of the zero-phonon line of SiV centers [2]. Thus, the diamond particle acts simultaneously as a heater and a thermometer. In the current work, we demonstrate the ability of such a Diamond Heater-Thermometer (DHT) to locally alter the temperature, one of the numerous parameters that play a decisive role for the living organisms at the nanoscale. In particular, we show that the local heating of 11-12 °C relative to the ambient temperature (22 °C) next to individual HeLa cells and neurons, isolated from the mouse hippocampus, leads to a change in the intracellular distribution of the concentration of free calcium ions. For individual HeLa cells, a long-term (about 30 s) increase in the integral intensity of Fluo-4 NW fluorescence by about three times is observed, which characterizes an increase in the $[Ca^{2+}]_{cyt}$ concentration of free calcium in the cytoplasm. Heating near mouse hippocampal neurons also caused a calcium surge - an increase in the intensity of Fluo-4 NW fluorescence by 30% and a duration of ~0.4 ms.

**<u>Introduction</u>**

Despite the constantly growing interest in local thermodynamic phenomena at micro and nanoscale in living cells and organisms in the last three decades [3]–[18], living cell thermodynamics is still a matter of guesses and controversies [19]–[22].

Discoveries of the faster neurite growth aligned with a local thermal gradient [11], heat-pulse induced muscle contractions without $Ca^{2+}$ influx [8] and highly thermosensitive $Ca^{2+}$ dynamics in HeLa cells through IP3 receptors [7] put forward the idea that endogenous intracellular heat generation events could be useful for the concept of "thermal signaling" [17]. This suggests that a living cell relies on ultralocal temperature variations of endogenous origin for the regulation of multiple physiological processes.

As mentioned in Ref. [17], future progress in studying intracellular thermodynamics needs an instrumentation revolution allowing local control of thermogenesis at micro/nanoscale, control of heat dissipation and heat energy conversion to other electrochemical energy types.

Currently, the development of new instruments for exploring intracellular thermodynamics mostly follows the idea of constructing Heater/Thermometer (HT) nanohybrids, combining nanomaterials able to generate heat and detect temperature. Multiple systems, such as metallic nanoparticles [23] and polymer coatings [24], efficiently transforming laser illumination into heat, have been reported as promising nanoscale heaters. Other materials, such as rare-earth complexes [25], quantum dots [26] and diamond nanocrystals [27], have been used to measure local temperature changes through the temperature-induced shift of their luminescence. However, such a hybrid HT approach complicates its practical use as one has to reproducibly match two nanostructures with different properties to form a hybrid one. Main disadvantages of the existing hybrid HTs include (=1) - the need to use a large number of hybrid nanoparticles in one experiment, (=2) - uncontrolled variations in heating properties from particle to particle, (=3) - the impossibility of preliminary calibration of an individual HT in terms of heating and temperature sensing, (=4) - the sensitivity to parameters of the external environment, in particular, pH.

We propose a design of a nanoscale heater/thermometer that is free from the disadvantages listed above. The device is based on the combination of the properties of a heater and a thermometer in one material - a polycrystalline diamond nanoparticle, and the use of only one HT, pre-calibrated in terms of temperature and heating, intended for experiments on ultralocal thermal effects in single cells and subcellular structures.

To demonstrate the capabilities of the developed diamond heater-thermometer (DHT), we present the results on thermal induction of intracellular calcium events in single cells (HeLa and neurons).

# Experimental details

**DHT design and operation**

The design principle of the DHT device follows that of the diamond thermometer described in detail in [2]. It consists of a polycrystalline CVD diamond particle embedded at the end facet of a submicron-diameter glass capillary (inset Fig. 1). The operating principle of the device is based on the dependence of the position of the maximum of the zero-phonon line (ZPL) luminescence of the "silicon-vacancy" (SiV) centers in the diamond particle on temperature [2], as well as on the ability of a diamond particle to absorb laser radiation due to the presence of amorphous carbon at its intercrystalline boundaries.

Before experimental use, a double calibration of the diamond particle is performed: via controlled temperature and via laser heating. The first calibration curve represents the dependence of the spectral position of the maximum of the zero-phonon luminescence line of SiV centers on a given temperature (see Methods). To perform the second calibration, a diamond particle fixed in a glass capillary is placed in the medium under study and the dependence of the heating temperature of the diamond nanoparticle on the power of the laser radiation focused on the particle is determined (see Results). The temperature calibration of a DHT was carried out in the low-power illumination regime ($P_{laser}<0.5$ mW), when the laser-induced heating of the particle is negligible (<0.1 °C). In contrast, the heating calibration was carried out by sweeping the laser power 1–5 mW range.

Therefore, a DHT is capable of operating in two modes. At higher laser power, it operates as the local heater and thermometer simultaneously. At low laser excitation (<0.5 mW) this device does not produce heating and operates as a thermometer.

**Experimental setup**

The study of the effect of local heating on living cells was carried out using a home-built confocal luminescent spectrometer (**Fig. 1**). Two lasers were used in the experiment: one to heat up the diamond particle, and the second one to excite the luminescence of the Fluo-4 NW dye used to visualize $[Ca^{2+}]_{cyt}$ in cells. The diamond particle was heated by a PicoQuant laser source at a wavelength of 561 nm, which was directed to the input of a water immersion objective focusing it on the DHT from the bottom of the coverslip. The DHT itself was fixed on a highly sensitive 3D micromanipulator, to enable its precise positioning in the focal spot of the objective, producing local heat release near the object of interest.

The second laser line at a wavelength of 473 nm (LaserComb) was illuminating the coverslip surface from the top through a long working distance objective (Mitutoyo, 20x). This illumination excited the fluorescence of the dye in the cells that were in the field of view of the objective (~40x40 um$^2$ area). The produced fluorescence near the surface of a selected living object was collected from the bottom of the coverslip through the water-immersion objective (Nikon 60x, WI, NA=1.27) and detected by a recording system consisting of an Andor Neo sCMOS camera and an Ocean Optics QEPro spectrometer. To filter out the spectral lines of laser sources, we used a combination of two optical filters LP 505 nm and SP 550 nm, effectively collecting the dye fluorescence from 505 to 550 nm (**Fig. 2**).

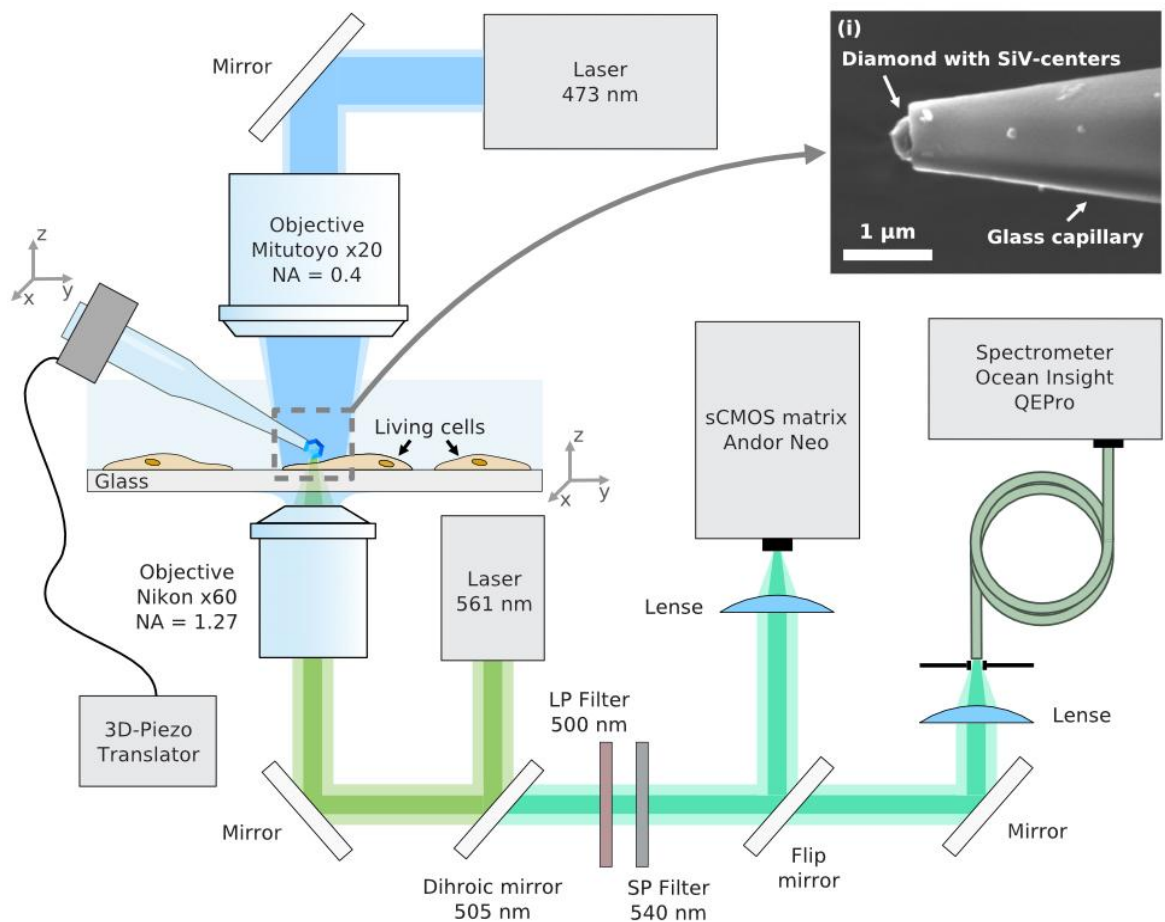

**Figure 1.** Schematic of the experimental setup for measuring the fluorescent response of the Fluo-4 NW dye in cells, and for tracking the luminescence of SiV centers in diamond. Two independent moving platforms are used: the coverslip seeded with living cells is fixed at the 3D-piezo stage, while the DHT is attached to the three-axis micromanipulator. The 473 nm laser beam illuminates the top side of the living cells through the low-NA objective. The excited Fluo-4 NW fluorescence collected from the bottom of the coverslip through the water-immersion objective (Nikon 60x, WI, NA=1.27) is then either projected on an sCMOS camera or sent to the spectrometer, depending on the position of the flip mirror. The 561 nm laser was used to excite the SiV-luminescence and to induce the heating; the inset (i) shows the SEM image of a submicron glass capillary with the diamond nanoparticle localized at the tip.

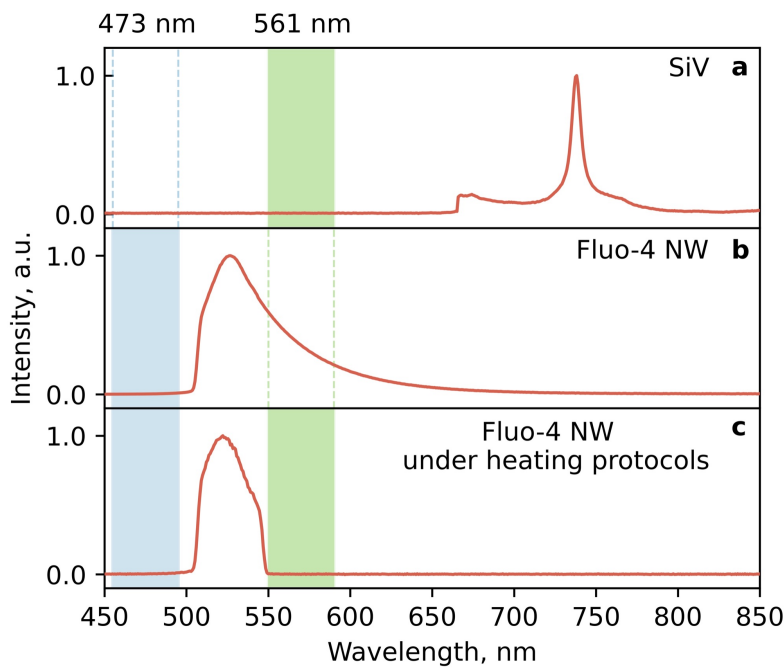

**Figure 2.** Characteristic experimentally-recorded emission spectra of: (a) the SiV-luminescence with 561-excitation, (b) Fluo-4 NW without the DHT with 473-excitation and (c) Fluo-4 NW following the DHT-heating protocol utilizing both 473 and 561 nm excitation lines simultaneously. The fluorescence signal is filtered out using a combination of short pass and long pass filters in the 505-550 nm range.

## Results

The temperature dependence of the calcium dynamics in the cytoplasm of living cells $[Ca^{2+}]_{cyt}$ has been studied using the cultured HeLa cells and primary neurons, isolated from the mouse hippocampus. To visualize the $[Ca^{2+}]_{cyt}$ we used the commercially-available dye (Fluo-4 NW, Thermofisher), following the protocol from the manufacturer. Briefly, the cell medium in the well plate was replaced by a 1 ml of a buffer solution (1X HBSS, 20 mM HEPES) mixed with the Fluo-4 NW dye and probenecid.

The diamond heater was preliminarily calibrated using the recorded heating temperature dependence on the illumination power at 561 nm. To do this, we have replaced the filter set from **Fig. 2c,** by another one transmitting long-wave radiation above 700 nm from SiV centers (see **Fig. 2a**). The DHT was placed in a buffer solution away from living objects and the SiV luminescence spectra were measured and the temperature was subsequently determined with an integration time of 1 s and in steps of 1 s for 100 seconds (**Fig. 3**). The resulting dependence shows three thermal steps with amplitudes of 25, 28.3, and 33.9 °C, depending on the 561 nm laser power. Finally, the last value was chosen for working with the cells. We used 22 °C bath temperature given the fact that intracellular calcium dynamics is well presented in the whole range of temperatures from 22 to 37 °C as shown in Ref. [7].

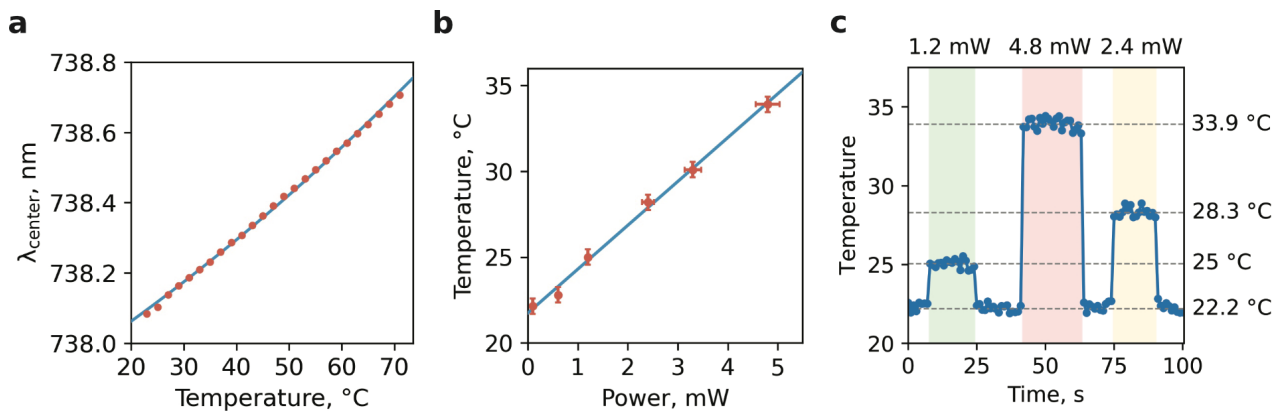

**Figure 3.** (a) - The calibration dependence of the ZPL spectral position from the temperature in the thermostat; (b) - The heating temperature calibration dependence on illumination power at 561 nm; (c) - Temporal changes in the heating temperature of a diamond particle at three different values of the illumination laser power (upper part of the graph). At a temperature of 22.2°C, the measured laser power before the objective was 0.2 mW; therefore, the baseline of the dependence (lower dashed line) characterizes the unperturbed system. An increase in illumination power leads to a proportional increase in temperature, which reaches its maximum Tmax = 33.9 °C at 4.8 mW.

At the next stage, the previously calibrated DHT was positioned next to a desired HeLa cell and fixed at a distance of ~0.2 µm from its surface. Then, at random moments, the 561 nm laser with a power of 4.8 mW was turned on for a short time (2-10 s) heating the diamond to 33.9 °C. Modeling the spatial temperature distribution of a spherical heat source in water [1] gives the maximum temperature close to the cell surface ≈ 29 oC. This temperature can be considered as instantly stabilized during the pulsed laser exposure (2-10 s), since the relaxation time to local thermal equilibrium in micron-scale aqueous medium is in the sub-millisecond range [31, 5], resulting in stepwise temperature clamp close to the cell surface. A steep temperature gradient near the nanoscale heat source in the water [1] results in ~1 μm$^2$ effective heated area of the living cell under our experimental conditions.

**Fig. 4** demonstrated the temporal response of the integrated fluorescence intensity of the dye F/F$_0$, where F/F$_0$ is determined as (F$_{high}$−F$_{low}$)/F$_{low}$. Here, F$_{high}$ is the average of fluorescence intensities for 1 s after it reaches the maximum during Ca$^{2+}$ overshoot, and F$_{low}$ is the average of consecutive three to five data points immediately after heat pulse, obtained for several HeLa cells during heat stimulation. The distribution of free intracellular calcium changes abruptly after the end of the heat stimulation (marked in pink). Following the cooling phase, one of the cells (**Fig. 4a**) exhibits a prolonged (~30 s) increase in [Ca$^{2+}$]$_{cyt}$. (3x magnification F/F$_0$). After the completion of the temperature disturbance for 90 s and 140 s, short calcium bursts with a duration of less than one second and an amplitude (F/F$_0$) of ~40% were detected. On the tail of the dependence, starting from t=150 s, the [Ca$^{2+}$]$_{cyt}$ level stabilizes, returning to the base level preceding the start of thermal stimulation. A similar increase in the intracellular level of free calcium immediately after the cooling phase is also observed for another cell (**Fig. 4b**). Here after 5 s of thermal exposure for 4 s, the fluorescence intensity (F/F$_0$) increases by 12%. **Figure 4c** shows a quasi-periodic temperature change (three protocols of stepwise heating of different duration), which allows one to effectively modulate in time the increase in the free concentration of calcium ions that occurs after each individual cooling phase of the temperature disturbance.

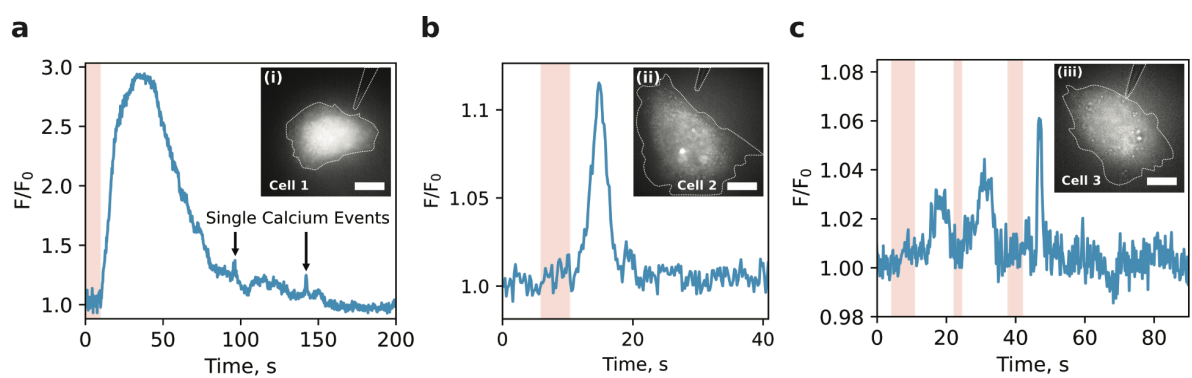

**Figure 4.** Time response of the normalized integrated fluorescence of Fluo-4 NW for three HeLa cells. The background component due to the presence of free calcium (plus the background noise of the recording system) in the buffer solution is experimentally measured beforehand and taken into account during normalization. The highlighted pink region indicates the time interval during which the heating was performed. The time resolution of each measurement is 0.2 s. Insets (i), (ii) and (iii) are the respective fluorescence images of HeLa cells before thermal stimulation. The dashed white lines prompt the outlines of cells and DHT. Scale bar: 10 µm.

We also performed a similar series of measurements with a primary culture of neurons isolated from the mouse hippocampus see **Figure 5.** The neuron staining procedure was performed according to the protocol (Thermo Fisher Scientific, Inc., Waltham, MA, USA). In individual neurons, an extremely fast response to a step change in temperature was observed, which consisted in an increase in the normalized integral intensity of Fluo-4 NW $[Ca^{2+}]_{cyt}$ fluorescence by ~30% within 0.4 s. As in the case of HeLa cells, the general trend revealed was an increase in the intracellular concentration of free calcium ions after each individual cooling phase.

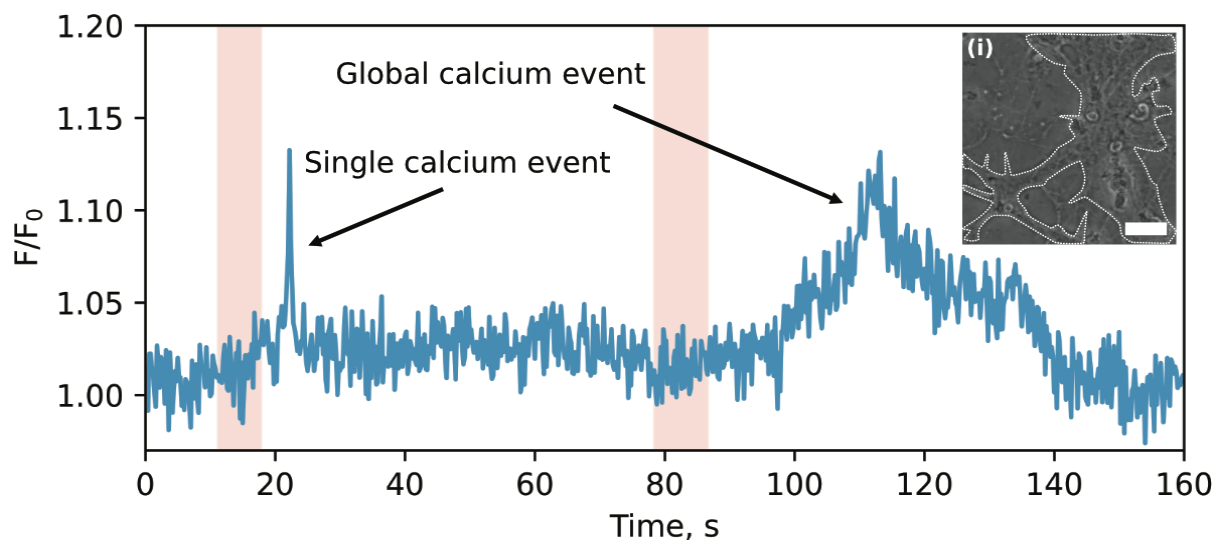

**Figure 5.** Time response of the normalized integrated fluorescence of Fluo-4 NW for a single neuron. The highlighted pink region indicates the time interval within which the heating was performed. The time resolution of each measurement is 0.2 s. The left calcium peak can be attributed to a rapid calcium spike, whereas the right one resembles a calcium wave since the cooling phase is followed by a steady calcium rise with time to peak at around half a minute. Inset (i) is the bright field image of the neuron before thermal stimulation. The dashed white lines prompt the outlines of cells. Scale bar: 10 µm.

## Discussion

In this work, we presented a new all-optical method for thermal stimulation of a single living cell with an accuracy of 0.1 °C [2] and a submicron spatial resolution, based on a DHT device. The main advantage of thermal stimulation with a DHT for a single living cell and its compartments is that the local temperature (temperature profile) is controlled reliably at a pre-defined point [2]. Moreover, the physical principles used to guarantee the absence of photobleaching in the DHT, enabling the use of unlimited experimental protocols for thermal stimulation of a living cell, which is impossible when using fluorescent temperature-sensitive sensors [4]–[6]. Combined with the biocompatibility of the DHT device, this creates a new method of stimulating a living cell - the Nanoscale Temperature Control Method, providing a great flexibility of thermal stimulation in comparison with the existing techniques.

As a particular example of the effectiveness of ultralocal thermal stimulation, we demonstrated a change in the level of free intracellular calcium in individual HeLa cells and in the primary culture of neurons isolated from the mouse hippocampus. We have revealed the general effect of an increase in the level of intracellular concentration of free calcium after the end of the stepwise temperature perturbation of a single living cell. Whereas all types of thermally induced calcium responses appeared next to the stepwise temperature downfall phase, we monitor significant differences in time duration (time to peak) of thermally induced calcium elevations. These time durations range from fraction to tens of seconds suggesting superposition of multifarious calcium related intracellular processes [28]. This is similar to the effect reported in the literature, utilizing local stepwise heating of aluminum nanoparticles with a 1064 nm laser and a fluorescent EuTTA thermometer [7]. However, our approach unifies the heater and thermometer in a single spot along with robust photostability allows the temperature protocols to be infinite in time, strongly spatially localized and highly effective. Calcium is a critical intra/extracellular signaling agent [28], so a new practical way to thermally manipulate and control the intracellular free calcium levels represents a major potential advancement in cell physiology in comparison with existing techniques such as, for example, flash photolysis of caged $Ca^{2+}$ [29].

In this regard, the thermodynamics of intracellular phenomena associated with calcium events itself would require new methods. For instance, it has been shown that the functioning of calcium pumps (CaATPase) in the endo-sarcoplasmic reticulum of a living cell leads to significant thermogenesis [30] as the concentration gradient built by the pump increases. Cyclic calcium release in tuna fish muscle-derived heater organs (and in other big fishes) is the main suspect responsible for significant thermogenesis capable of keeping the eyes and brain of fish at 15 °C higher temperature in cold water[31]. Such data do not indirectly exclude the possible connection and synergy of universal intracellular calcium signaling with the thermal signaling concept [17].

Therefore, the fundamental importance of the new way of ultralocal temperature manipulation constitutes in the fact that for the first time it practically introduces this parameter into cellular physiology, namely into the privileged order of operational ultralocal parameters of the electrochemical potential equation[32], [33] similarly to the transmembrane electric potential or ionic concentrations. This new tool is compatible with Patch Clamp [34] and $Ca^{2+}$ imaging [35] and finally tames the temperature at the nanoscale – the last parameter of the equation of electrochemical potential still escaping its precise ultralocal experimental control in intracellular compartments.

## Methods

### DHT calibration

The developed DHT was placed in an air environment for a preliminary determination of the power density of optical excitation at a wavelength that did not lead to a shift of the zero-phonon SiV luminescence line, and, consequently, to heating. The measured power density was used to determine the temperature dependence of the spectral position of the zero-phonon line center in a Linkam TS1500 thermostat, stabilized at a predetermined level with an accuracy of ~1 °C. The temperature in the chamber was changed in increments of 10 °C. At each step, the SiV luminescence spectrum was recorded and the spectral position of the zero-phonon line maximum was determined according to the algorithm proposed in our previous work [2].

### Cell culture

Cell culture HeLa cells were cultured in Dulbecco's modified eagle's medium (DMEM) (Invitrogen, CA, USA) supplemented with fetal bovine serum (10%) and penicillin–streptomycin. The HeLa cell suspension was uniformly dispersed in a 2-inch Petri dish filled with 1 ml of the cell medium (DMEM) and cultured for 20 hours. Cells were grown in a glass-based dish at 37 °C in the presence of 5% $CO_2$.

### Primary neuron culture preparation

Pregnant female C57BL/6 J Rcc Hsd were obtained from Harlan Laboratories (France) and were housed according to the Swiss legislation and the European Community Council directive (86/609/EEC). Primary hippocampal cultures were prepared from mice brains from P0 pups as previously described [36]. The protocol was approved by the Swiss cantonal authorities VD, 'Service de la consommation et des affaires vétérinaires' Lausanne, Switzerland. Briefly, the hippocampi were isolated stereoscopically and dissociated by trituration in a medium containing papain (20 U ml$^{-1}$, Sigma-Aldrich, Switzerland). The neurons were plated on FluoroDish Sterile Culture Dish 35 mm, 23 mm well (World Precision Instruments), previously coated with poly-L-lysine 0.1% w/v in water, and neurons were cultured in Neurobasal medium containing B27 supplement (Life Technologies), L-glutamine and penicillin/streptomycin (100 U ml$^{-1}$, Life Technologies).

### Acknowledgements

This work was supported by the Human Frontier Science Program RGP0047/2018 (to Vadim Zeeb). Evgenii Glushkov and Aleksandra Radenovic acknowledge the support from National Center of Competence in Research (NCCR) Bio-Inspired Materials and Max-Planck-EPFL Center of Molecular Nanoscience and Technology. We would like to thank prof. Hilal Lashuel's laboratory (Laboratory of Chemical Biology of Neurodegeneration, EPFL), especially (Dr. Nathan Riguet and Ms.Yllza Jasiqi), for providing the neuronal cells. Also, we would like to thank Dr. Rodrigo Perin from the laboratory of prof. Henry Markram (Laboratory of Neural Microcircuitry,EPFL) for lending us the micromanipulator to control the DHT.

### Data availability

All data generated or analysed during this study are included in this published article.